\documentclass[twocolumn,conference]{IEEEtran}
\usepackage[T1]{fontenc}
\usepackage{float}
\usepackage{amsmath}
\usepackage{amsthm}
\usepackage{graphicx}

\makeatletter

\newcommand{\noun}[1]{\textsc{#1}}
\providecommand{\tabularnewline}{\\}
\floatstyle{ruled}
\newfloat{algorithm}{tbp}{loa}
\providecommand{\algorithmname}{Algorithm}
\floatname{algorithm}{\protect\algorithmname}

  \theoremstyle{definition}
  \newtheorem{example}{\protect\examplename}
  \theoremstyle{definition}
  \newtheorem{defn}{\protect\definitionname}
  \theoremstyle{definition}
  \newtheorem{problem}{\protect\problemname}
  \theoremstyle{plain}
  \newtheorem{thm}{\protect\theoremname}
  \theoremstyle{plain}
  \newtheorem{lem}{\protect\lemmaname}

\usepackage[caption=false,font=footnotesize]{subfig}

\makeatother

\providecommand{\definitionname}{Definition}
\providecommand{\examplename}{Example}
\providecommand{\lemmaname}{Lemma}
\providecommand{\problemname}{Problem}
\providecommand{\theoremname}{Theorem}

\begin{document}

\title{Diversified Top-$k$ Similarity Search in Large Attributed Networks}

\author{{\normalsize{}Zaiqiao Meng$^{a}$ and Hong Shen$^{a,b}$}\\
\textit{\normalsize{}$^{\text{a}}$School of Information Science and
Technology, Sun Yat-Sen University, China}\\
\textit{\normalsize{}$^{\text{b}}$School of Computer Science, University
of Adelaide, Australia}\\
\textit{\normalsize{}Email: zqmeng@aliyun.com, hongsh01@gmail.com}{\normalsize{} }}
\maketitle
\begin{abstract}
Given a large network and a query node, finding its top-\emph{k} similar
nodes is a primitive operation in many graph-based applications. Recently
enhancing search results with diversification have received much attention.
In this paper, we explore an novel problem of searching for top-\emph{k}
diversified similar nodes in attributed networks, with the motivation
that modeling diversification in an attributed network should consider
both the emergence of network links and the attribute features of
nodes such as user profile information. We formulate this practical
problem as two optimization problems: the \emph{Attributed Coverage
Diversification (ACD)} problem and the \emph{r-Dissimilar Attributed
Coverage Diversification (r-DACD)} problem. Based on the submodularity
and the monotonicity of \emph{ACD}, we propose an efficient greedy
algorithm achieving a tight approximation guarantee of $1-1/e$. Unlike
the expension based methods only considering nodes' neighborhood,
\emph{ACD} generalize the definition of diversification to nodes'
own features. To capture diversification in topological structure
of networks, the \emph{r-DACD} problem introduce a dissimilarity constraint.
We refer to this problem as the \emph{Dissimilarity Constrained Non-monotone
Submodular Maximization} (DCNSM) problem. We prove that there is no
constant-factor approximation for DCNSM, and also present an efficient
greedy algorithms achieving $1/\rho$ approximation, where $\rho\le\Delta$,
$\Delta$ is the maximum degree of its dissimilarity based graph.
To the best of our knowledge, it is the first approximation algorithm
for the \emph{Submodular Maximization} problem with a distance constraint.
The experimental results on real-world attributed network datasets
demonstrate the effectiveness of our methods, and confirm that adding
dissimilarity constraint can significantly enhance the performance
of diversification.
\end{abstract}

\section{Introduction\label{sec:Introduction}}

Searching for top-$k$ nodes similar to a given query request in a
network has numerous applications including graph clustering \cite{huang2015dense},
graph query \cite{mottin2015graph}, and object retrieval and recommendation
\cite{wang2015friendbook}. There has been substantial research on
ranking nodes and estimating the similarity (proximity) between nodes,
such as the Personalized PageRank \cite{haveliwala2002topic} and
SimRank\cite{jeh2002simrank}. These basic methods and their variations
such as P-Rank \cite{zhao2009p}, TopSim \cite{lee2012top} and Panther
\cite{zhang2015panther} has been successfully applied in many applications.

Nowadays with rich information available from online social networks,
real social entities and their relationships can be built in a network
in which nodes are associated with a set of attributes describing
their properties and edges represent relationships among these nodes.
In such circumstances, the problem of searching similar nodes for
a given node becomes more sophisticated and challenging. 

Firstly, the top-$k$ nodes resulted from the traditional similarity
searching methods are often highly related. It is hard to give a glimpse
of the overall similar results with such few and highly related nodes.
Search result diversification has been widely studied as a way of
tackling query ambiguity and enhancing result novelty in information
retrieval \cite{carbonell1998use,gollapudi2009axiomatic}. Most of
these diversified models try to trade-off the relevance for diversity
in the results, thereby making the result list more diverse. In the
literature there are many studies on modeling the search result diversification
for network datasets, based on node's ego features diversification,
such as neighborhood, exemplified by the recently proposed models
of expansion ratio \cite{li2013scalable} and expanded relevance \cite{kuccuktuncc2013diversified}
applying the solution to the classic \emph{Cardinality Constrained
Monotone Submodular Maximization} problem. However, A major drawback
of these models is that they have no explicit measure to eliminate
redundancy resulting in a lack of novelty in their search results. 

Moreover, unlike simplified structural networks only with nodes and
edges, a network with attributes has more complex characters. Node
attributes along with the links between them provide rich and complementary
sources of information and should be used simultaneously for uncovering,
understanding and exploiting the latent diversified structure in attributed
network data. For example, in social networks users have profile information,
and in document networks each node also contains the text of the document
that it represents, hence diversification is also presence among these
non-topological information. Above-mentioned diversification models
can find diverse nodes which account for the topological connections
between nodes, but they cannot account for the node attributes. 

In this paper, to modeling the diversification search problem in attributed
networks, we first formulate the problem that only considers the \emph{attribute
coverage diversification} (ACD). We prove that the optimization objective
of the ACD problem is a nondecreasing submodular function, and a marginal
gain based greedy algorithm can obtains an $(1-1/e)$-approximation
near-optimal solution. To to improve novelty, we add an $r$-dissimilar
constraint to ACD problem for capturing the dissimilarity between
result nodes based on the topological structure of the graph. We show
that the new problem become more complicated, because adding a new
node to the result set following the monotonicity may break the dissimilarity
constraint, the existing techniques can not be applied to our dissimilarity
constrained diversification model. Our model requires to solve a new
problem of\emph{ Dissimilarity Constrained Non-monotone Submodular
Maximization} (DCNSM). Based on constructing a dissimilarity-based
graph, we propose a greedy algorithm achieving an approximation ratio
of $1/\rho$ , where $\rho$ is bounded by the maximum degree $\Delta$
of its dissimilarity-based graph, and runs in $O((\Delta+k)\left(|V||A|+|E|\right))$
time.

The main contributions of this paper are: (1) We formalize the dissimilarity
constrained diversification model for top-$k$ diversified similarity
search in attributed network that combining attribute coverage diversification
with a dissimilarity constraint as the problem of maximizing a dissimilarity
constrained non-monotone submodular function problem, we prove that
there is no constant-factor approximation for this problem and present
a linear time algorithm with approximation ratio $1/\rho$, where
$\rho$ is bounded by the maximum degree of its dissimilarity-based
graph. (2) We conduct extensive experiments on real-world attributed
network datasets, the results shows the effectiveness of our proposed
algorithms; we also combined the baseline models with the dissimilarity
constraint, and also shows that the new dissimilarity constrained
methods significantly outperforms the original models.

\section{Problem Formulation\label{sec:problem}}

\subsection{Preliminaries}

Let $G=(V,E,W)$ be an undirected weighted graph with $|V|$ nodes
and $|E|$ edges. $W$ is the \emph{edge} weighting function such
that $W=\left\{ w:E(G)\rightarrow R+\right\} $. If the weighting
function is not specified, the weight of each edge is $1$. We denote
$V(G)$ as the set of nodes in $G$, $G[V]$ denote the subgraph of
$G$ induced by $V$. Let $d_{G}(u)$ be the degree of node $u\in V$,
$\Delta$ the maximum degree of $G$, $N_{G}(u)$ the neighborhood
of $v$, and $N_{G}^{+}(u)$, $\left\{ u\right\} \bigcup N_{G}(u)$. 

Given an undirected weighted network $G$, a positive integer $k$,
and a query node $q\in V$, the relevance metric $s(\cdot)$ is the
similarity score of each nodes measuring the relevance to $q$. Analogously,
a dissimilarity metric $diss(\cdot,\cdot)$ is the dissimilarity score
measuring the dissimilarity between pair of nodes. In this paper,
the similarity metric and the dissimilarity metric are defined and
computed purely based on the topological structural context of nodes.

Given an element $u$, a ground set $N$, a function $f:2^{N}\rightarrow R^{+}$
is called \emph{submodular function} if for $\nabla S\subseteq T\subseteq N$
and $u\in N\backslash B$, $f_{u}(S)\ge f_{u}(T)$, which $f_{u}(S)=f(S+u)-f(S)$
is called the \emph{marginal gain} . A submodular function $f$ is
\emph{monotone} if for every $S\subseteq T$ we have that $f(S)\leq f(T)$
\cite{nemhauser1978analysis}.

\subsection{Problem Definition}

\begin{figure}[tbh]
\includegraphics[width=8cm]{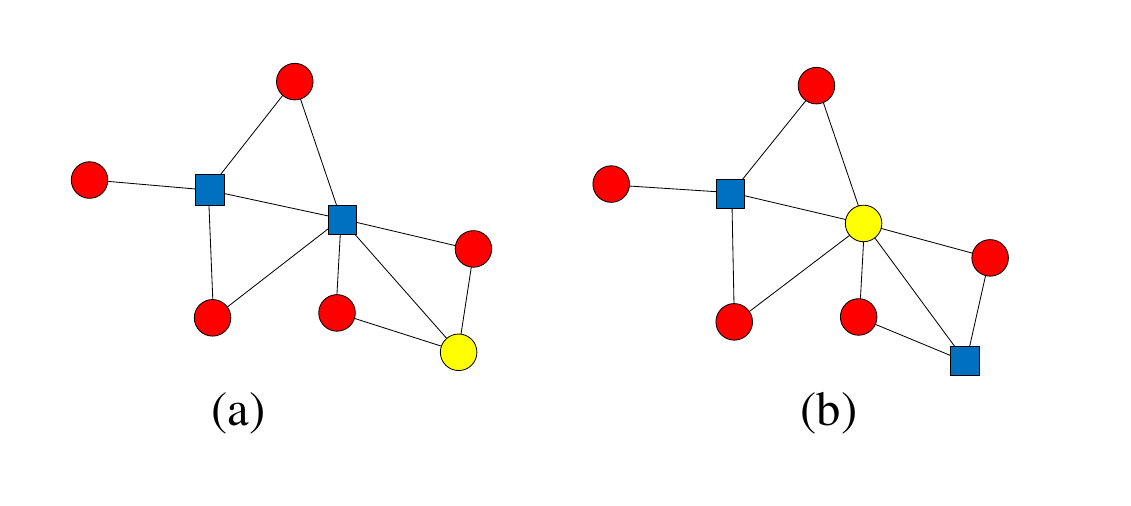}

\caption{\label{fig:Exp-based}(a) Diversified result from EP1 (b) A more novelty
diversified result}
\end{figure}

We target the problem of diversifying top-$k$ similarity search result
for a given node in networks, assuming that the relevance score for
each node the has already obtained by a given similarity search algorithms
which we'll talk about more. The main challenge of this problem is
how to properly measure the diversification. Some previous models
try to optimize a function with a single diversification measure based
on the ego features of nodes, such as neighborhood. For example, \cite{li2013scalable}
formulate this problem as a bicriteria objective optimization problem
that tradeoff the relevance and the\emph{ expansion ratio}, \cite{kuccuktuncc2013diversified}
optimize a single function (called \emph{expanded relevance}) which
combines both relevance and neighborhood diversity. These neighbor
expansion based methods are based on the intuition that nodes with
large expansion are dissimilar to each other, thus leading to diversity,
and they don't have a specific metric to measure the dissimilarity
ensuring the novelty. However, result nodes with large expansion set
may not always indicate that they are dissimilar to each other. We
illustrate this in Example \ref{example:exp}.
\begin{example}
\label{example:exp}Consider a graph in Fig. \ref{fig:Exp-based},
and suppose all the nodes has a same relevance score. The blue square
nodes in Fig. \ref{fig:Exp-based} (a) and (b) are the possible result
returned by EP1 algorithm \cite{li2013scalable} and the circle nodes
denote their expansion nodes, because the expansion ratio of the selected
nodes in both Fig. \ref{fig:Exp-based} (a) and Fig. \ref{fig:Exp-based}
(b) is 1. But it is clearly that result nodes in Fig. \ref{fig:Exp-based}
(b) are more novelty to each other than the result nodes in Fig. \ref{fig:Exp-based}
(a). Because the selected nodes in Fig. \ref{fig:Exp-based} (b) are
more dissimilar to each other whether it is measured based on shortest
path or based on common neighbors. 
\end{example}

Thus, the neighbor expansion based methods does not guarantee novelty
of their result sets. Moreover, all existing algorithms do not consider
the node attributes when it is involved in an attributed network. 

To effectively perform diversified search, we must first determine
what the diversified results is. A good diversification metric should
eliminate the redundant nodes and choose a set of representative nodes
which are dissimilar to each other. In real-world social networks
with rich categorical node attributes, diversified search should consider
both network structure and node attribute information. Our main task
of this paper is to model the diversified search problem in attributed
networks, considering the both network structure and node attribute
information, and trading-off the diversity and the relevance. 

Based on the undirected weighted graph, we give the definition of
the attributed network in our problem setting. 
\begin{defn}
\noun{Undirected Weighted Attributed Network.} Let $G=(V,E,W_{E},W_{A})$
denotes an \emph{undirected weighted attributed network}, where\emph{
}$W_{E}$ is the \emph{edge} weighting function such that $W_{E}=\left\{ w:E(G)\rightarrow R+\right\} $,
$W_{A}$ is the \emph{node} attributed-weighting function such that
$W_{A}=\left\{ w:E(G)\rightarrow S_{A}\right\} $, where $S_{A}$
is a family subsets of attribute set $A$, which representing categorical
attributes of each nodes. 
\end{defn}
We assume that the attribute of nodes are binary-valued. Other types
of attribute variables could be clustered into categorical variables
via vector quantization, or discretized to categorical variables.
For example, in Facebook social network, various universities (e.g.,
MIT, CMU, and Stanford) in user profile are directly treated as separate
binary attributes, the posted status updates can use stemming, tokenization,
and stopword removal etc. techniques to extract keywords as binary
attribute. Some reduction and classification technique in attribute
could be used to further improve the performance but is not the scope
of this paper. 

To model the diversification in attributed networks, we first introduce
the \emph{attribute coverage diversification} problem which only considering
the diversification on attribute features of node, then we propose
a new model of diversification combining diversification of node attributes
with a dissimilarity constrain such that any two nodes in the search
result must satisfy a given dissimilarity threshold to ensure the
novelty, and cover attributes as much as possible. 
\begin{defn}
\label{def:AC} \noun{Attribute Coverage}. Let $S$ be a set of nodes,
the \emph{attribute coverage} of node $v$, $|A_{v}|$, is the cardinality
of attribute set of $v$, $|A_{s}|=|\cup A_{v}|$ is the \emph{attribute
coverage} of node subset $S$. The \emph{attribute coverage ratio}
(ACR) of $S$, $ACR=\frac{|A_{s}|}{|A|}$, is defined as the normalization
of attribute coverage, where $|A|$ is the total number of attribute
set. \end{defn}
\begin{problem}
\label{prob:ACD}\noun{Attribute Covering Diversification (ACD)}.
Given an undirected weighted attributed network $G=(V,E,W_{E},W_{A})$,
a query node $q$, a relevant metric $s(\cdot)$, and a positive integer
$k$, the problem is to find a subset $S\subseteq V$ that:
\begin{eqnarray}
 & \underset{S\subseteq V}{\max}~f(S)=(1-\lambda)\underset{u\in S}{\sum}s(u)+\lambda\frac{|A_{S}|}{|A|}\nonumber \\
 & s.t.~~|S|=k.~\;~~~~~~~~~~~\;\;~~~~~~~~~~\;\;\;\;\;~
\end{eqnarray}

\end{problem}
where $\lambda\in[0,1]$ is a parameter to tradeoff relevance and
diversity, $\frac{|A_{s}|}{|A|}$ is the \emph{attribute coverage
ratio} representing the attributed diversity. Problem \ref{prob:ACD}
utilizes the linear combination of the attribute coverage and the
relevant metric as the optimization objective, and aim at finding
a subset $S$ of $k$ nodes such that: (1) the nodes in $S$ have
high relevance to the query node $q$; (2) the result subset $S$
have maximum attribute coverage. The definition of attribute diversity
base on attribute coverage is intuitive and reasonable, it indicate
that the more attributes covered by nodes, the more diverse the result
set is. However, the ACD problem only considers one diversification
measure that is the attribute diversification, and still can not guarantee
novelty of their result sets. To consider the diversification on network
topology and improve the novelty of result, the dissimilarity in topological
structure between nodes is taken into account in this problem. Formally,
we then model the new dissimilarity-constrained diversified search
problem. 
\begin{problem}
\noun{\label{prob:rDACD}$r$-Dissimilar Attribute Coverage Diversification
($r$-DACD)}. Given an undirected weighted attributed network $G=(V,E,W_{E},W_{A})$,
a query node $q$, a relevant metric $s(\cdot)$, a dissimilarity
metric $diss(\cdot,\cdot)$, and a positive integer $k$, the problem
is to find a subset $S\subseteq V$ that: 
\begin{eqnarray}
 & \underset{S\subseteq V}{\max}~f(S)=\underset{u\in S}{(1-\lambda)\sum}s(u)+\lambda\frac{|A_{S}|}{|A|}\nonumber \\
 & s.t.~~|S|=k,~\;~~~~~~~~~~~\;~~~~~~~~~~\;\;\;~\;\;\\
 & ~~\;\;\;\;\;\;\;\forall v_{1},v_{2}\text{\ensuremath{\in}}S,~~diss(v_{1},v_{2})\leq r.~~\;\;~\nonumber 
\end{eqnarray}

\end{problem}
In this problem, the $r$-dissimilar constraint scatters the result
to a wider range of topological structure space, describes the topological
structural diversification of the result set. The relevant measurement
and dissimilar metrics in Problem \ref{prob:ACD} and Problem \ref{prob:rDACD}
will be discussed in the next section.

\subsection{Connection to Neighbor Expansion based Methods}

There exists many other previous work studies the bicriteria optimization
measures in diversified search on graphs problems. For example, \cite{li2013scalable}
tries to optimize the following diversified bicriteria optimization
objective: 
\begin{equation}
\underset{S\subseteq V}{\max}f(S)=(1-\lambda)\sum_{u\in S}s(u)+\lambda\frac{|N_{S}|}{|N|}.
\end{equation}

The first term is the sum of the Personalized PageRank scores over
the results, which reflects the relevance. The second term is the
expansion ratio of the results, which reflects the diversity.

We notice that our ACD problem is the ego feature generalization of
this neighbor expansion diversification problem. If we view the neighbor
sets of nodes as their attributed sets, the neighbor expansion based
diversified problem will become our ACD problem. In other words, the
neighbors is the attributes of nodes, to some degree. 

Another neighbor expansion based measure called \emph{expanded relevance}
which combines both relevance and diversity into a single function
in order to measure the coverage of the relevant part of the graph
\cite{kuccuktuncc2013diversified}. They argue that bicriteria optimization
is inappropriate, because the diversification methods that seem to
optimize both criteria are highly correlated among each other. It
is worth mentioning that both ACD and $r$-DACD also optimize a bicriteria
objective, the structural similarity and the attributed diversity,
but they obviously are two independent metrics to measure relevance
and diversity. To capture the novelty among result nodes, we also
add the same dissimilarity constraint to these neighbor expansion
based methods. We will illustrate the comparison of these dissimilarity
constrained models in detail in the follow-up experiments.

\section{Algorithms\label{sec:algorithms}}

\subsection{Relevance Metric and Dissimilarity Metric}

Given a large graph, finding top-$k$ most similarity (proximity)
nodes to a given query node is a fundamental problem. Most of previous
work about diversified search on graphs utilize Personalized PageRank
(PPR) \cite{haveliwala2002topic} as their relevance metric. In this
paper we use Panther \cite{zhang2015panther} to measure the relevance
to query node and the dissimilarity between pair of nodes in result
set, the major reasons are as follows. Firstly, compared with PPR,
Panther has a lower time complexity which is unrelated to the node
size, so that it can deal with large-scale networks. Secondly, we
need an unified measure to accommodate both the relevant and the dissimilarity
of nodes, and using Panther can easily meet this needs. 

The basic idea of Panther is that two nodes are similar if they frequently
appear on the same paths. The algorithm randomly select a vertex in
$G$ as the starting point, and then conduct random walks of $T$
steps from $v$ using the weight proportion as the transition probability
to corresponding nodes. The relevance score is defined as: $s(u)=\frac{|p_{u,q}|}{R}$,
where $|p_{u,q}|$ is the number of paths that contains node $u$
and the query node $q$, and $R$ is the total number of random paths.
The authors show that Panther can provably and accurately estimates
the similarity between any pair of nodes \cite{zhang2015panther}.

In this paper, we also evaluate the dissimilarity of nodes by the
idea that two nodes are dissimilar if they don't occurs frequently
together on the same paths. The normalized dissimilar score is defined
as: $diss(u,v)=1-\frac{|p_{u,v}|}{|p_{max}|-|p_{min}|}$, where $|p_{max}|$,
$|p_{min}|$ represent the maximum and minimum number of random paths
between two nodes, and $|p_{u,v}|$ is the number of random paths
between node $u$ and node $v$.

\subsection{The Greedy ACD Algorithm }

It is easy to see that the ACD problem is NP-hard, because if we let
$\lambda=1$, the problem is equivalent to the \emph{max k-cover}
problem which is known to be NP-hard \cite{feige1998threshold}. Thus
we resort to develop approximate algorithms for this problem. Below,
we prove that Problem \ref{prob:ACD} is a nondecreasing submodular
function with a cardinality constraint \cite{buchbinder2014submodular}.
\begin{thm}
\label{thm:CCMSM}The optimization problem of $f(S)$ defined in Problem
\ref{prob:ACD} is a Cardinality Constrained Monotone Submodular Maximization
problem. 
\end{thm}
Theorem \ref{thm:CCMSM} can be proved by the same way as the proof
of Theorem 3.2 in \cite{li2013scalable}, so we omit the proof here.

Although maximization of submodular functions is NP-hard, there exists
efficient algorithms with theoretical approximation guarantee for
this problem. We propose to construct a greedy solution that repeatedly
chooses a node with the \emph{maximum marginal gain} until the size
of result set equal to $k$. Algorithm \ref{alg:GACD} outline the
\emph{greedy ACD} algorithm (GACD).

\begin{algorithm}[tbh]
\caption{\label{alg:GACD}Greedy ACD}

\textbf{Input:} An undirected weighted attributed network $G$, $k$,
$\lambda$, query node $q$, relevance metric $s(\cdot)$ to $q$

\textbf{Output:} A set $S$ with $k$ nodes

1: $i\leftarrow0$, $S\leftarrow\emptyset$, $G_{i}=G$;

2: \textbf{while} $|S|<k$ \textbf{do}

3: $\quad$update $f_{u}(S)$ for nodes in $G_{i}$;

4: $\quad$$v\leftarrow\max_{u\in V\left(G_{i}\right)}f_{u}\left(S\right)$;

5: $\quad$$S=S\bigcup\left\{ v\right\} $; 

6: $\quad$$G_{i+1}=G[V\left(G_{i}\right)-v]$;

7: $\quad$$i=i+1$; 

8: \textbf{return} $S$; 
\end{algorithm}

In this algorithm, the most of time cost is computing the marginal
gain $f_{u}(S)$ for every nodes (step 3). We can use an indicator
array to represent the attribute coverage, then the time complexity
from step 3 to step 4 is $O(|V||A|)$. Thus, the\emph{ }total time
complexity is $O(k|V||A|)$.
\begin{thm}
Algorithm \ref{alg:GACD} is an $(1-1/e)$-approximation algorithm
for Problem \ref{prob:ACD}.
\end{thm}
According to the results in \cite{nemhauser1978analysis}, for a monotone
submodular function, greedily constructing by selecting an element
with the maximum marginal gain gives an $(1-1/e)$-approximation to
the optimal. Since from Theorem \ref{thm:CCMSM}, we can easily prove
that Algorithm \ref{alg:GACD} is an $(1-1/e)$-approximation algorithm
(We omit the proof here).

\subsection{The Greedy $r$-DACD Algorithm}

The main difficulty in designing algorithm for $r$-DACD problem is
that considering the dissimilarity constraint makes the optimization
objective become non-monotone, because adding a new node to the result
set may break the dissimilarity constraint. The greedy algorithm used
to solve the \emph{monotone submodular maximization} problem can not
directly be deployed for $r$-DACD problem. As far as we know, there
were no reports and literature of the relevant studies mention this
dissimilarity constrained problem. We denote this problem as \emph{Dissimilarity
(Distance) Constrained} \emph{Non-monotone Submodular Maximization
}(DCNSM) problem. In fact, the follow theorem shows that the DCNSM
problem is NP-hard even hard to approximate within any constant factor.
\begin{thm}
\label{thm:hard}It is hard to approximate for the DCNSM problem within
$n^{1-\epsilon}$, for any $\epsilon$>0, unless $NP=ZPP$. \end{thm}
\begin{IEEEproof}
We show that DCNSM problem is reducible to the maximum$\mathbf{}$
independent set (MIS) problem via a costpreserving reduction; since
MIS strictly equivalent to the maximum clique problem, which is hard
to approximate within $n^{1-\epsilon}$, for any $\epsilon>0$, unless
$NP=ZPP$ \cite{haastad1996clique}. 

Given a submodular function $f$, a ground set $N$, a dissimilarity
function $diss(v_{1},v_{2})$ for every pair of elements $v_{1},v_{2}\in N$.
To prove the theorem, we construct a new dissimilarity-based graph
$G^{'}=(V^{'},E^{'},f)$ defined as follows: for any element $u\in N$,
$u$ is a node of $V^{'}$ in $G^{'}$, for any pair of element that
violate the dissimilarity constraint ($diss(v_{1},v_{2})<r$), $(v_{1},v_{2})$
is an edge between $v_{1},v_{2}$ in $G^{'}$. $f$ is the node submodular
function such that for any subset $S\in V^{'}$, $f(S)$ satisfy the
submodularity. From this, construct a simplified instance of DCNSM
problem, we set the submodular function $f(v)=1$ for any single element,
$f(S)=|S|$ for any subset $S\subseteq V^{'}$. Clearly, the construction
of $G^{'}$ can be done in polynomial time. We see that any solution
of MIS in $G^{'}$ is precisely the subset $I\subseteq V^{'}$ maximizing
$f(I)$ and satisfying the dissimilarity constraint, and vice versa. 

This completes the proof. 
\end{IEEEproof}
Clearly, the DCNSM problem with cardinality constraint also satisfy
Theorem \ref{thm:hard}. To solve the $r$-DACD problem is to solve
cardinality constrained DCNSM problem, we put forward the \emph{Greedy
$r$-DACD} algorithm (G$r$DACD). Details of the algorithm are presented
in Algorithm \ref{alg:GrDACD}.

\begin{algorithm}[tbh]
\textbf{Input:} An undirected weighted attributed network $G$, $k$,
$\lambda$, query node $q$, relevance metric $s(\cdot)$ to $q$,
a dissimilarity metric $diss(\cdot,\cdot)$

\textbf{Output:} A set $S$ with $k$ nodes

1: generate dissimilarity-based graph $G^{'}$;

2: $\rho\leftarrow1$,$i\leftarrow0$, $S\leftarrow\emptyset$, $G_{i}^{'}=G^{'}$;

3: \textbf{while} $|S|<k$ \textbf{do} 

4: $\quad$update $f_{u}(S)$ for nodes in $G_{i}^{'}$;

5: $\quad$$v\leftarrow\max_{u\in V\left(G_{i}^{'}\right)}f_{u}\left(S\right)$,

$\quad$$\quad$$s.t.$ $f_{u}(S)\ge\frac{1}{\rho}\sum_{w\in N_{G_{i}^{'}}(u)}f_{w}(S)$; 

6: $\quad$\textbf{if} $v_{i}$ is $NULL$ \textbf{then} 

7: $\quad$$\quad$$\rho=\rho+1$; \textbf{continue}; 

8: $\quad$$S=S\bigcup\left\{ v\right\} $; 

9: $\quad$$G_{i+1}^{'}=G_{i}^{'}[V\left(G_{i}^{'}\right)-N_{G_{i}^{'}}^{+}\left(v\right)]$;

10: $\quad$$i=i+1$; 

11:\textbf{ return} $S$; 

\caption{\label{alg:GrDACD}Greedy $r$-DACD }
\end{algorithm}

In Algorithm \ref{alg:GrDACD}, the algorithm first constructing a
dissimilarity-based graph $G^{'}$, then greedily select a node with
the\emph{ }maximum marginal gain that is greater than the $1/\rho$
times of the summed marginal gain of its neighbors, where $\rho$
is denoted as \emph{local maximal factor}. In each iteration, the
algorithm compute the marginal gain $f_{u}(S)$ for each $u\in V\left(G_{i}^{'}\right)$,
and chooses a node $u$ with maximum $f_{u}(S)$ and satisfying the
condition: $f_{u}(S)\ge\frac{1}{\rho}\sum_{w\in N_{G_{i}^{'}}(u)}f_{w}(S)$,
what we call as the\emph{ local maximal node}. The algorithm first
set $\rho=1$, if there is no local maximal node, then let $\rho=\rho+1$,
and continue for the next iteration. We show that the algorithm can
always return $k$ local maximal nodes within $\rho\le\Delta$ , where
$\Delta$ is the maximum degree of $G^{'}$.
\begin{lem}
\label{lem:bound}Algorithm \ref{alg:GrDACD} can terminate when $\rho\le\Delta$. \end{lem}
\begin{IEEEproof}
To prove the lemma, we prove that there always exists at least one
local maximal node in $G_{i}^{'}$ when $\rho=\Delta$. Assume that
$\rho=\Delta$, and there is no local maximal node, then for each
$u\in G_{i}^{'}$ the following inequality holds:

$f_{u}(S)<\frac{1}{\Delta}\sum_{w\in N_{G_{i}^{'}}(u)}f_{w}(S)$

Accumulating this inequality for each nodes in $G_{i}^{'}$, we can
get:

$\sum_{u\in G_{i}^{'}}f_{u}(S)<\sum_{u\in G_{i}^{'}}\frac{1}{\Delta}\sum_{w\in N_{G_{i}^{'}}(u)}f_{w}(S)$

The number of times $f_{w}(S)$ appeared on the right side exactly
is the degree of $w$, because each $f_{w}(S)$ on the right side
represents that there exists an edge between $u$ and $w$ in $G_{i}^{'}$.
Thus:

$\sum_{u\in G_{i}^{'}}f_{u}(S)<\sum_{w\in G_{i}^{'}}\frac{d(w)}{\Delta}f_{w}(S)$

Rearranging yields:

$\sum_{u\in G_{i}^{'}}\frac{\Delta}{\Delta}f_{u}(S)<\sum_{u\in G_{i}^{'}}\frac{d(u)}{\Delta}f_{u}(S)$

From $d(u)\leq\Delta$, we reach a contradiction, hence $\rho\le\Delta$.\end{IEEEproof}
\begin{thm}
\label{thm:mswisapp}Algorithm \ref{alg:GrDACD} gives a $\frac{1}{\rho}$-approximation
for Problem \ref{prob:rDACD}, where $\rho\le\Delta$, $\Delta$ is
the maximum degree of its dissimilarity-based graph . \end{thm}
\begin{IEEEproof}
We just need to prove it based on the approximate solution of the
generated dissimilarity-based graph $G^{'}$. Let $OPT(G^{'})$ be
optimal value for $G^{'}$, $v_{i}\in S$ be the selected node with
the order $i$ by Algorithm\textit{ }\ref{alg:GrDACD}. The algorithm
starts with $S=\left\{ \emptyset\right\} $, and ends when $|S|=k$
with an optimal value of  ${\displaystyle \sum_{i=1}^{k}}f_{v_{i}}\left(S\right)$. 

Now we consider the induced subgraph $G^{'}\left(N_{G_{i}^{'}}^{+}(v_{i})\right)$
. The optimal solution of $G^{'}\left(N_{G_{i}^{'}}^{+}(v_{i})\right)$
can be $v_{i}$ or a subset of $N_{G_{i}^{'}}(v_{i})$, and the optimal
value satisfies:

$OPT\left(G^{'}\left(N_{G_{i}^{'}}^{+}(v_{i})\right)\right)\leq\max\left(f_{v_{i}}(S),f_{N_{G_{i}^{'}}(v_{i})}(S)\right)$

From the local maximal condition, we have:

$f_{v_{i}}(S)\geq\frac{1}{\rho}\sum_{w\in N_{G_{i}^{'}}(v_{i})}f_{w}(S)$

Combining the two inequation we have:

$f_{v_{i}}(S)\geq\frac{1}{\rho}OPT\left(G^{'}\left(N_{G_{i}^{'}}^{+}(v_{i})\right)\right)$

For each $v_{i}\in S$, their the induced subgraph $N_{G_{i}^{'}}^{+}(v_{i})$
is independent, according to the submodularity of $f$, the following
inequality holds:

${\displaystyle \sum_{i=1}^{k}}OPT\left(G\left(N_{G_{i}^{'}}^{+}(v_{i})\right)\right)\geq OPT\left(G^{'}\right)$

Then we have:

${\displaystyle \sum_{i=1}^{k}}f_{v_{i}}\left(S\right)\geq{\displaystyle \sum_{i=1}^{k}}\frac{1}{\rho}OPT\left(G^{'}\left(N_{G_{i}^{'}}^{+}(v_{i})\right)\right)\geq\frac{OPT\left(G^{'}\right)}{\rho}$

From Lemma \ref{lem:bound}, the upper bound of $\rho$ is $\Delta$,
then we complete the proof.
\end{IEEEproof}
Now we analysis the time complexity of Algorithm\textit{ }\ref{alg:GrDACD}.
For each loop, computing the marginal gain $f_{u}(S)$ for every nodes
in $G_{i}$ (step 4) needs $O(|V||A|)$ time. Finding the \emph{local
maximal node} in step 6 will takes $O(|E|)$ time. The loop will be
terminated in at most $\Delta+k$ iterations, hence the complexity
of G$r$ACD is $O((\Delta+k)\left(|V||A|+|E|\right))$, which is linear
w.r.t. the node size, the edge size and the attribute size.

Although G$r$DACD does not achieve any constant-factor approximation,
its implements on the real-world data to conduct the experiment can
achieve good results.

\section{\label{sec:setup}Experimental Setup}

\subsection{\label{sub:datasets}Datasets}

For our experiments, we consider three real-world datasets where we
have network topological information as well as node attributes. The
brief statistical information of our datasets are presented in Table
\ref{tab:Sum_Dataset}.

The Facebook dataset is downloaded from SNAP\footnote{http://snap.stanford.edu/data/}.
This dataset is built from profile and relation data from 10 users'
ego-networks in Facebook, and the attributes are constructed by their
user profiles. The DBLP dataset is from the DBLP public bibliography
data\footnote{http://dblp.uni-trier.de/xml/}. We build a coauthor
network extracting from the papers in top 172 conferences (rank A
and B) from 10 research areas ranked and classified by CCF\footnote{http://www.ccf.org.cn/sites/paiming/2015ccfmulu.pdf}.
We treat each author as a vertex, each collaboration of paper as an
edge. The 172 conferences are viewed as the attributes of each node.
The AMiner coauthor dataset is downloaded from AMiner.org\footnote{https://aminer.org/AMinerNetwork}.
This network dataset is also built by collaboration relationships
among the authors, but unlike the DBLP dataset, the attributes in
AMiner dataset are constructed by the extracted keyterms of their
papers. 

\begin{table}[tbh]
\caption{\label{tab:Sum_Dataset}Summary of the real-world datasets}

\centering{}%
\begin{tabular}{|c|c|c|c|}
\hline 
datasets & nodes & edges & attributes\tabularnewline
\hline 
\hline 
Facebook & 4,039 & 88,234 & 1,406\tabularnewline
\hline 
DBLP & 73,242  & 373,797 & 172\tabularnewline
\hline 
AMiner & 1,560,640 & 4,258,946 & 2,868,034\tabularnewline
\hline 
\end{tabular}
\end{table}

\subsection{\label{sub:metrics}Evaluation Metrics }

In the literature, there are no well accepted measures for diversified
search in graph, since it is different to the \textquotedblleft diversification\textquotedblright{}
definition. In our experiments, we first employ two common metrics
appeared in references to measure the relevance and the diversity
in topological structure. The first one is the \emph{normalized relevance}
($Rel$) which is given in \cite{tong2011diversified}, defined as
$Rel=\frac{\sum_{u\in S}s(u)}{\sum_{u\in\tilde{S}}s(u)}$, where $S$
denotes the top-$k$ diversified result list by the diversified algorithms,
$\tilde{S}$ denotes the top-$k$ similarity nodes list by relevance
metric. By definition, the normalized relevance measures the similarity
of result to the query request, and the higher $Rel$ implies the
better relevance. The second metric is the \emph{density} of the induced
subgraph of the result set. The density of a graph is the number of
edges existing in the graph divided by the maximum possible number
of edges in the graph. In the topological structure perspective, the
less density implies the more diverse of the result.

We also introduce two new metrics for our new problems: the \emph{attributed
coverage ratio} (ACR) and the \emph{minimum dissimilarity} ($MinDiss$).
The attributed coverage ratio is defined in Definition \ref{def:AC},
which measure the attributed diversity. The average dissimilarity
is defined as $MinDiss=Min_{v_{1},v_{2}\text{\ensuremath{\in}}S}diss(v_{1},v_{2})$,
which measure the structural diversity.

\subsection{\label{sub:baselines}Baselines}

We compare our proposed methods with several state-of-the-art baselines:
the bicriteria expansion optimization \cite{li2013scalable} (denoted
by EP) and the expanded relevance \cite{kuccuktuncc2013diversified}
(denoted by BC). For our experiments, we mainly focus on $l$-step,
for $l=1$ and $l=2$, denoted by EP1, EP2 and BC1, BC2 respectively.
As EP1, EP2, BC1 and BC2 also can be build on the dissimilarity constraint
version, we also consider four \textbf{\textit{$r$}}-Dissimilar constraint
variants of these methods, denoted by $r$-DEP1, $r$-DEP2 and $r$-DBC1,
$r$-DBC2 respectively.

\subsection{\label{sub:environment}Parameter Settings and Experimental Environment}

For the experiments based on diversified search problems, we fist
run the Panther algorithm to obtain the relevance score for all nodes,
extract the candidate nodes that have some relevance to the query
node and compute the dissimilarity between all pairs of candidate
nodes, then run the diversified algorithms to obtain their diversified
search results. There are two parameters in Panther: path length $T$
and error-bound $\epsilon$. We empirically set $T=5$ and $\epsilon\approx\sqrt{1/|E|}$
which can obtain a good performance on all the datasets \cite{zhang2015panther}.
In GACD, G$r$DACD, EP1, EP2, $r$-DEP1 and $r$-DEP2 algorithms,
the candidate set size is 2000, because the maximum result set size
$k$ of our experiments is 100, and nodes with relevance rank more
than 2000 are often irrelevant to the query node. In BC1, BC2, $r$-DBC1
and $r$-DBC2, the candidate nodes are those nodes with a relevance
score of more than 0.0001. In diversified search algorithms, there
are two parameters: $\lambda$ used to tradeoff relevance and diversity,
and $r$ used to restrict the dissimilarity of results. We use the
different parameter $\lambda$ value in different algorithms, because
the scale coefficient of their diversity function is different. 

All experiments were conducted on an Ubuntu 14.04 server with two
Intel Xeon E5-2683 v3 (2.0GHz) CPU and 128G RAM. All algorithms are
implemented by C++.

\section{\label{sec:results}Results and Analysis}

\subsection{\label{sub:meticscomparision} Comparison of Metrics}

\begin{figure*}[tbh]
\begin{centering}
\includegraphics[width=18cm]{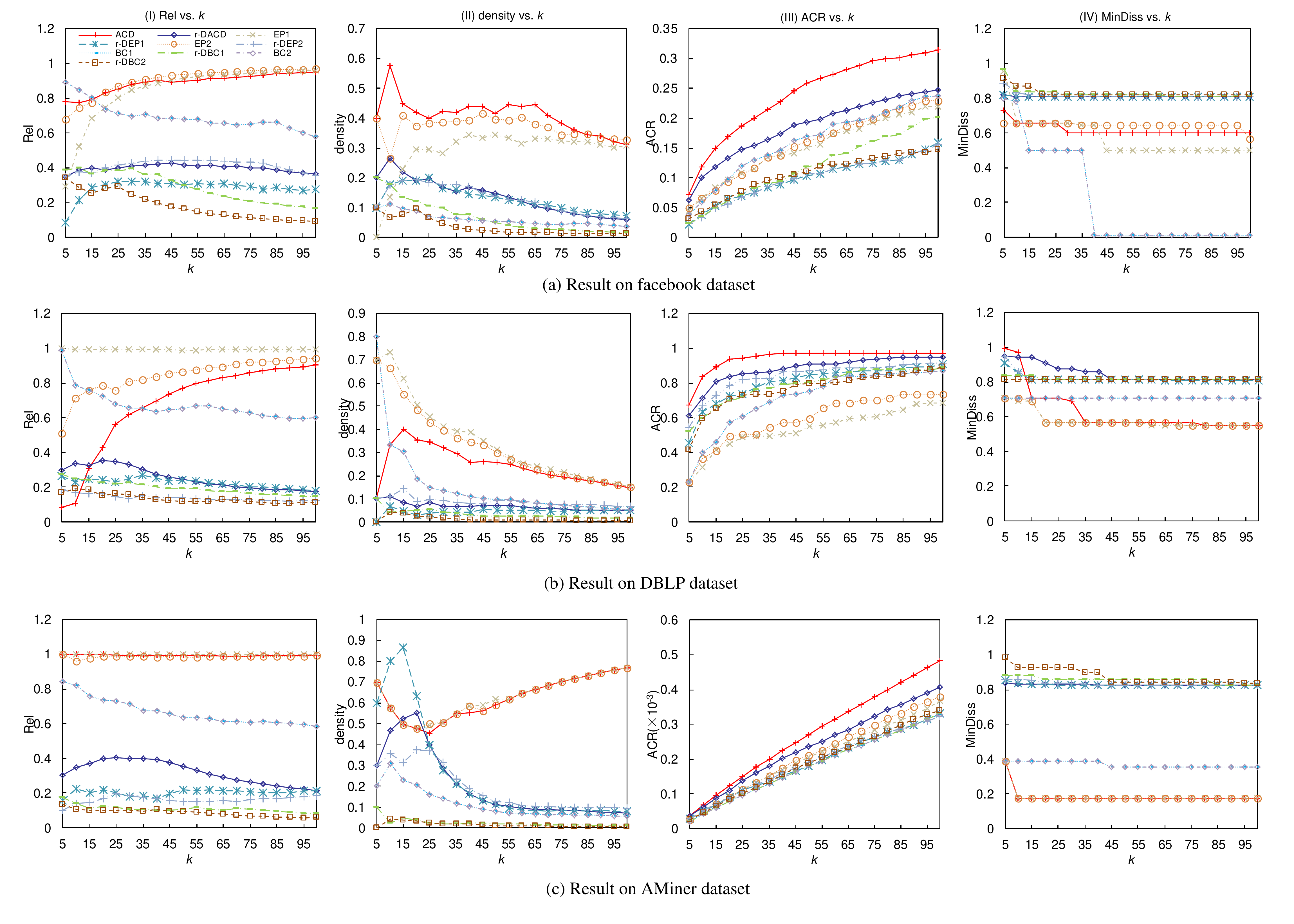}
\par\end{centering}

\caption{\label{fig:metricscom}Comparison under various diversity and relevance
metrics }
\end{figure*}

In this subsection, we evaluate the effectiveness of GACD and G$r$DACD
under diversity and relevance metrics defined above. We run all the
algorithms on the three real-world network datasets with varying $k$
values ($k\in\left[5,100\right]$). Normalized relevance ($Rel$)
plots in the first column of Fig. \ref{fig:metricscom} shows that
GACD achieve comparable score in Facebook and AMiner datasets, but
are worse than EP1 and EP2 in DBLP dataset. We can also observe the
fact in the first column of Fig. 2 that, when we combine the dissimilarity
constraint, $Rel$ decreased significantly in all the datasets, and
our G$r$DACD can keep a relatively high $Rel$ score. Although a
low $Rel$ score is not an indication of being dissimilar to the query,
but excessive high $Rel$ score usually implies that the algorithm
ignored the diversity of the results. 

We also plot the density results in the second column of Fig. \ref{fig:metricscom}
and observe a similar result. When we add a dissimilarity constraint
to these methods, the density decreased significantly compared with
GACD and original baselines, and $r$-DBC gets the lowest density
in all the datasets. Recall that the lower density indicates the less
similar to each other, which represents the more diverse in structural
topology to some extent. However, the normalized relevance and the
density can not measure the diversity in attributed coverage. 

Hence, we measure the ACR score of these returned results. Experimental
comparison in the third column of Fig. \ref{fig:metricscom} shows
that our proposed GACD outperforms the competitors (recall that the
higher value of ACR, the more diverse attributes are), and G$r$DACD
also achieve a good score of ACR only behind GACD in all datasets.
Certainly, the reason behind this result is that our GACD and G$r$DACD
assign the attribute covering ratio as the optimizing objective, while
others do not consider attributes which leading to relatively poor
results. Based on the results on real-world network datasets, we conclude
that our GACD and G$r$DACD can effectively obtain the attributed
diversified search result, and the G$r$DACD exhibit both better attributed
diversity and better structural topology diversity. 

To further validate the effectiveness of the dissimilarity constraint,
we evaluate the $MinDiss$ performence for all the algorithms. The
results are depicted in the fourth column of Fig. \ref{fig:metricscom}.
In this set of figures, we can clearly see that results from the dissimilarity
constrained algorithms all have a higher $MinDiss$ score than GACD
and the original of neighbor expansion based algorithms. And the score
of BC1 and BC2 are even near zero in Facebook network, which means
that there exist at least two nodes that are extremely similar to
each other in the result set. We can say that those result sets are
lack of novelty, and adding the dissimilarity constraint can significantly
enhance the novelty.

\subsection{\label{sub:Scalability}Scalability}

\begin{figure*}[tbh]
\begin{centering}
\textsf{\includegraphics[width=13.5cm]{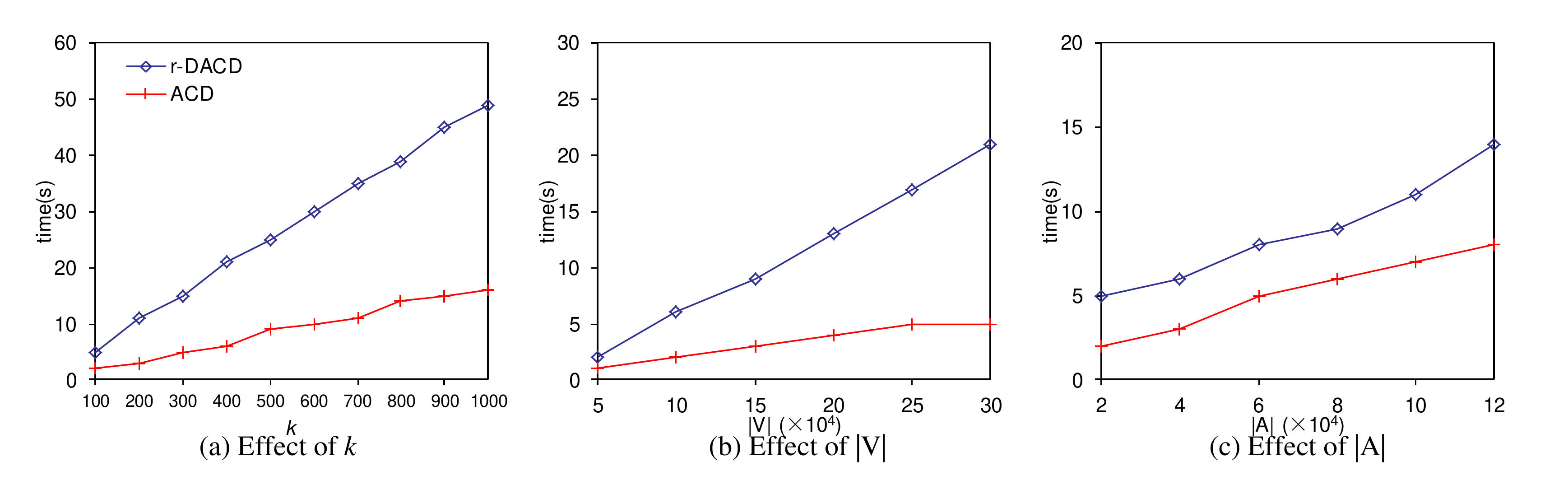}}
\par\end{centering}

\caption{\label{fig:Scalability}Scalability of our algorithms}
\end{figure*}

To evaluate the scalability performance of our proposed algorithms,
we execute the two algorithms on a series of synthetic datasets generated
by Erdös-Rényi (ER) random network model \cite{erdds1959random}  over
different $k$, with node size ranging from 50,000 to 300,000, total
set-weight length ranging from 2,0000 to 120,000. With the runtime
experiments shown in Fig. \ref{fig:Scalability}, we can clearly see
that both GACD and G$r$ACD scale linearly w.r.t the result size,
the number of nodes and the number of attributes, and the GACD algorithm
outperform slightly. This confirms our time complexity analysis in
the previous sections, thus they all can be scalable to large networks.

\subsection{\label{sub:Case}Case Study}

\begin{table*}[tbh]
\caption{\label{tab:Case-study}Case study}

\centering{}\includegraphics[width=16cm]{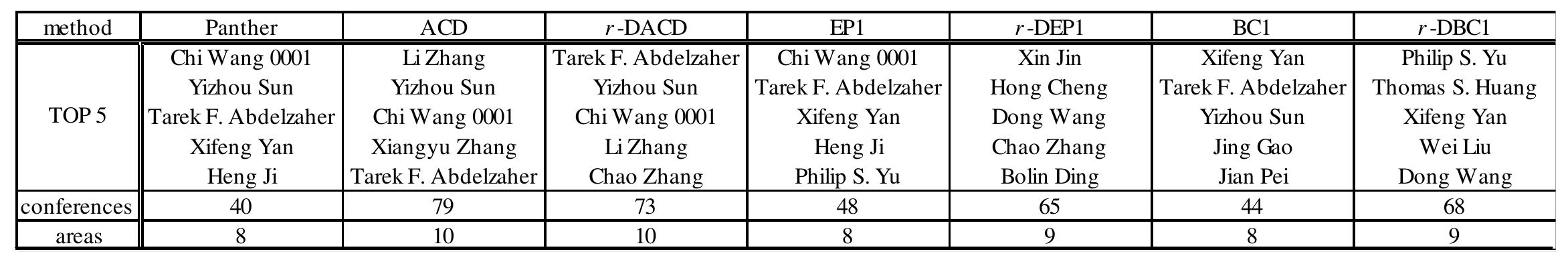}
\end{table*}

Now we present a case study to demonstrate the effectiveness of the
proposed methods. Table \ref{tab:Case-study} shows an example of
top-5 diversified search results for Jiawei Han in DBLP network. We
also count the number of conferences (attributes) that the top-5 authors
covered and the number of research areas that conferences belong to.
As shows in Table \ref{tab:Case-study}, the 7 methods present very
different results. Those authors found by Panther have closest relevance
with the query author, but have the worst coverage in conference and
research area. On the contrary, our GACD and G$r$DACD achieves the
best and the next best performance in both conference and research
area coverage. And both $r$-DEP1 and $r$-DBC1 also are significantly
better performance than EP1 and BC1. These results not only confirm
the effectiveness of the proposed our GACD and G$r$DACD methods,
and also confirm that combining dissimilarity constraint can enhance
the performance of diversification.

\section{Related Work\label{sec:related}}

There has been various measures to estimate similarity between nodes
on networks. Personalize PageRank (PPR) \cite{haveliwala2002topic}
is a random walk based measures evolved from the classic PageRank
algorithm \cite{page1999pagerank}. Similar to PPR, SimRank is defined
recursively with respect to the ``random surfer-pairs model'', it
evaluates the similarity between two nodes as the first-meeting probability
of two random surfers. Existing algorithms like P-Rank \cite{zhao2009p},
TopSim \cite{lee2012top} are the extension of SimRank. Some other
examples include discounted/truncated hitting time \cite{sarkar2010fast},
penalized hitting probability \cite{zhang2012evaluating}, and nearest
neighbor \cite{bogdanov2013accurate,wu2014fast} are also referred
the random walk method. Recently, a random path sampling based method---Panther
\cite{zhang2015panther} was proposed that can provably, fast and
accurately estimates the similarity between nodes.

Submodularity is a property of set functions with deep theoretical
consequences and far-reaching applications. Submodular set functions
has been widely applied to many fields, including document summarization
\cite{lin2010multi}, image segmentation \cite{jegelka2011submodularity},
sensor placement \cite{krause2008near}, diversifying search \cite{li2013scalable,borodin2012max},
and algorithmic game theory \cite{dughmi2009revenue}. Submodular
function maximization captures classic NP-hard problems in the combinational
optimization such as \emph{max cut} problems, \emph{maximum facility
location} problems and \emph{max k-cover} problems \cite{buchbinder2014submodular,buchbinder2015tight,feige2011maximizing,vondrak2013symmetry}.
Some research deals with maximizing submodular functions subject to
various combinatorial constraints, such as the bases of a matroid
\cite{vondrak2013symmetry}, multiple knapsack constraints \cite{kulik2009maximizing}
and submodular knapsack \cite{iyer2013submodular}. To the best of
our knowledge, there is no published work providing studies in the
problem of maximizing submodular functions subject to a dissimilarity
(distance) constraint.

There are several studies on search results diversification in network
data. DivRank\cite{mei2010divrank} employs a time-variant random
walk process to facilitates the rich-gets-richer mechanism in node
ranking. Tong, et al. \cite{tong2011diversified} propose a scalable
diversified ranking algorithm by optimizing a predefined diversified
goodness measure. Recently a neighbor expansion based diversified
ranking method was proposed, with the assumption that nodes with large
expansion would be dissimilar to each other \cite{li2013scalable}
. Küçüktunç \cite{kuccuktuncc2013diversified} propose a measure called
expanded relevance which combines both relevance and diversity into
a single function in order to measure the coverage of the relevant
part of the graph. These methods are designed to work with the simplified
structural network without considering node attributes. Our diversification
models focus on the attributed networks, and combines with a dissimilarity
constraint as an explicit measure to eliminate redundancy .

\section{\label{sec:conclusion}Conclusion}

In this paper, we explore a practical problem of diversifying search
results in attributed networks. Based on modeling attributed diversification
problem (ACD), we formulate this problem as the $r$-DACD problem
that combining attributed diversification with dissimilarity-constrained
diversification to improve novelty of search results. We show that
the $r$-DACD problem is hard to approximate within any constant factor.
Two approximation algorithms was proposed to solve these two problems
with bounded approximation ratio. We empirically compare our algorithms
with two state-of-the-art diversified search methods, as well as their
improved algorithms combining a dissimilarity constraint, in terms
of both the structural diversified metrics and the attributed diversified
metrics in real-world attributed networks. The results shows the effectiveness
of our proposed algorithms, and also confirms that combining dissimilarity
constraint in diversification can significantly improve the query
result on novelty.

\bibliographystyle{IEEEtranS}
\bibliography{Diversification}

\begin{thebibliography}{10}
\providecommand{\url}[1]{#1}
\csname url@samestyle\endcsname
\providecommand{\newblock}{\relax}
\providecommand{\bibinfo}[2]{#2}
\providecommand{\BIBentrySTDinterwordspacing}{\spaceskip=0pt\relax}
\providecommand{\BIBentryALTinterwordstretchfactor}{4}
\providecommand{\BIBentryALTinterwordspacing}{\spaceskip=\fontdimen2\font plus
\BIBentryALTinterwordstretchfactor\fontdimen3\font minus
  \fontdimen4\font\relax}
\providecommand{\BIBforeignlanguage}[2]{{%
\expandafter\ifx\csname l@#1\endcsname\relax
\typeout{** WARNING: IEEEtranS.bst: No hyphenation pattern has been}%
\typeout{** loaded for the language `#1'. Using the pattern for}%
\typeout{** the default language instead.}%
\else
\language=\csname l@#1\endcsname
\fi
#2}}
\providecommand{\BIBdecl}{\relax}
\BIBdecl

\bibitem{bogdanov2013accurate}
P.~Bogdanov and A.~Singh, ``Accurate and scalable nearest neighbors in large
  networks based on effective importance,'' in \emph{Proceedings of the 22nd
  ACM international conference on Conference on information \& knowledge
  management}.\hskip 1em plus 0.5em minus 0.4em\relax ACM, 2013, pp.
  1009--1018.

\bibitem{borodin2012max}
A.~Borodin, H.~C. Lee, and Y.~Ye, ``Max-sum diversification, monotone
  submodular functions and dynamic updates,'' in \emph{Proceedings of the 31st
  symposium on Principles of Database Systems}.\hskip 1em plus 0.5em minus
  0.4em\relax ACM, 2012, pp. 155--166.

\bibitem{buchbinder2014submodular}
N.~Buchbinder, M.~Feldman, J.~S. Naor, and R.~Schwartz, ``Submodular
  maximization with cardinality constraints,'' in \emph{Proceedings of the
  Twenty-Fifth Annual ACM-SIAM Symposium on Discrete Algorithms}.\hskip 1em
  plus 0.5em minus 0.4em\relax SIAM, 2014, pp. 1433--1452.

\bibitem{buchbinder2015tight}
N.~Buchbinder, M.~Feldman, J.~Seffi, and R.~Schwartz, ``A tight linear time
  (1/2)-approximation for unconstrained submodular maximization,'' \emph{SIAM
  Journal on Computing}, vol.~44, no.~5, pp. 1384--1402, 2015.

\bibitem{carbonell1998use}
J.~Carbonell and J.~Goldstein, ``The use of mmr, diversity-based reranking for
  reordering documents and producing summaries,'' in \emph{Proceedings of the
  21st annual international ACM SIGIR conference on Research and development in
  information retrieval}.\hskip 1em plus 0.5em minus 0.4em\relax ACM, 1998, pp.
  335--336.

\bibitem{dughmi2009revenue}
S.~Dughmi, T.~Roughgarden, and M.~Sundararajan, ``Revenue submodularity,'' in
  \emph{Proceedings of the 10th ACM conference on Electronic commerce}.\hskip
  1em plus 0.5em minus 0.4em\relax ACM, 2009, pp. 243--252.

\bibitem{erdds1959random}
P.~ERDdS and A.~R\&WI, ``On random graphs i,'' \emph{Publ. Math. Debrecen},
  vol.~6, pp. 290--297, 1959.

\bibitem{feige1998threshold}
U.~Feige, ``A threshold of ln n for approximating set cover,'' \emph{Journal of
  the ACM (JACM)}, vol.~45, no.~4, pp. 634--652, 1998.

\bibitem{feige2011maximizing}
U.~Feige, V.~S. Mirrokni, and J.~Vondrak, ``Maximizing non-monotone submodular
  functions,'' \emph{SIAM Journal on Computing}, vol.~40, no.~4, pp.
  1133--1153, 2011.

\bibitem{gollapudi2009axiomatic}
S.~Gollapudi and A.~Sharma, ``An axiomatic approach for result
  diversification,'' in \emph{Proceedings of the 18th international conference
  on World wide web}.\hskip 1em plus 0.5em minus 0.4em\relax ACM, 2009, pp.
  381--390.

\bibitem{haastad1996clique}
J.~H{\aa}stad, ``Clique is hard to approximate within $n^{1-\epsilon}$,'' in
  \emph{Foundations of Computer Science, 1996. Proceedings., 37th Annual
  Symposium on}.\hskip 1em plus 0.5em minus 0.4em\relax IEEE, 1996, pp.
  627--636.

\bibitem{haveliwala2002topic}
T.~H. Haveliwala, ``Topic-sensitive pagerank,'' in \emph{Proceedings of the
  11th international conference on World Wide Web}.\hskip 1em plus 0.5em minus
  0.4em\relax ACM, 2002, pp. 517--526.

\bibitem{huang2015dense}
X.~Huang, H.~Cheng, and J.~X. Yu, ``Dense community detection in multi-valued
  attributed networks,'' \emph{Information Sciences}, vol. 314, pp. 77--99,
  2015.

\bibitem{iyer2013submodular}
R.~K. Iyer and J.~A. Bilmes, ``Submodular optimization with submodular cover
  and submodular knapsack constraints,'' in \emph{Advances in Neural
  Information Processing Systems}, 2013, pp. 2436--2444.

\bibitem{jegelka2011submodularity}
S.~Jegelka and J.~Bilmes, ``Submodularity beyond submodular energies: coupling
  edges in graph cuts,'' in \emph{Computer Vision and Pattern Recognition
  (CVPR), 2011 IEEE Conference on}.\hskip 1em plus 0.5em minus 0.4em\relax
  IEEE, 2011, pp. 1897--1904.

\bibitem{jeh2002simrank}
G.~Jeh and J.~Widom, ``Simrank: a measure of structural-context similarity,''
  in \emph{Proceedings of the eighth ACM SIGKDD international conference on
  Knowledge discovery and data mining}.\hskip 1em plus 0.5em minus 0.4em\relax
  ACM, 2002, pp. 538--543.

\bibitem{krause2008near}
A.~Krause, A.~Singh, and C.~Guestrin, ``Near-optimal sensor placements in
  gaussian processes: Theory, efficient algorithms and empirical studies,''
  \emph{The Journal of Machine Learning Research}, vol.~9, pp. 235--284, 2008.

\bibitem{kuccuktuncc2013diversified}
O.~K{\"u}{\c{c}}{\"u}ktun{\c{c}}, E.~Saule, K.~Kaya, and {\"U}.~V.
  {\c{C}}ataly{\"u}rek, ``Diversified recommendation on graphs: pitfalls,
  measures, and algorithms,'' in \emph{Proceedings of the 22nd international
  conference on World Wide Web}.\hskip 1em plus 0.5em minus 0.4em\relax ACM,
  2013, pp. 715--726.

\bibitem{kulik2009maximizing}
A.~Kulik, H.~Shachnai, and T.~Tamir, ``Maximizing submodular set functions
  subject to multiple linear constraints,'' in \emph{Proceedings of the
  twentieth Annual ACM-SIAM Symposium on Discrete Algorithms}.\hskip 1em plus
  0.5em minus 0.4em\relax Society for Industrial and Applied Mathematics, 2009,
  pp. 545--554.

\bibitem{lee2012top}
P.~Lee, L.~V. Lakshmanan, and J.~X. Yu, ``On top-k structural similarity
  search,'' in \emph{Data Engineering (ICDE), 2012 IEEE 28th International
  Conference on}.\hskip 1em plus 0.5em minus 0.4em\relax IEEE, 2012, pp.
  774--785.

\bibitem{li2013scalable}
R.-H. Li and J.~X. Yu, ``Scalable diversified ranking on large graphs,''
  \emph{Knowledge and Data Engineering, IEEE Transactions on}, vol.~25, no.~9,
  pp. 2133--2146, 2013.

\bibitem{lin2010multi}
H.~Lin and J.~Bilmes, ``Multi-document summarization via budgeted maximization
  of submodular functions,'' in \emph{Human Language Technologies: The 2010
  Annual Conference of the North American Chapter of the Association for
  Computational Linguistics}.\hskip 1em plus 0.5em minus 0.4em\relax
  Association for Computational Linguistics, 2010, pp. 912--920.

\bibitem{mei2010divrank}
Q.~Mei, J.~Guo, and D.~Radev, ``Divrank: the interplay of prestige and
  diversity in information networks,'' in \emph{Proceedings of the 16th ACM
  SIGKDD international conference on Knowledge discovery and data
  mining}.\hskip 1em plus 0.5em minus 0.4em\relax Acm, 2010, pp. 1009--1018.

\bibitem{mottin2015graph}
D.~Mottin, F.~Bonchi, and F.~Gullo, ``Graph query reformulation with
  diversity,'' in \emph{Proceedings of the 21th ACM SIGKDD International
  Conference on Knowledge Discovery and Data Mining}.\hskip 1em plus 0.5em
  minus 0.4em\relax ACM, 2015, pp. 825--834.

\bibitem{nemhauser1978analysis}
G.~L. Nemhauser, L.~A. Wolsey, and M.~L. Fisher, ``An analysis of
  approximations for maximizing submodular set functions—i,''
  \emph{Mathematical Programming}, vol.~14, no.~1, pp. 265--294, 1978.

\bibitem{page1999pagerank}
L.~Page, S.~Brin, R.~Motwani, and T.~Winograd, ``The pagerank citation ranking:
  bringing order to the web.'' 1999.

\bibitem{sarkar2010fast}
P.~Sarkar and A.~W. Moore, ``Fast nearest-neighbor search in disk-resident
  graphs,'' in \emph{Proceedings of the 16th ACM SIGKDD international
  conference on Knowledge discovery and data mining}.\hskip 1em plus 0.5em
  minus 0.4em\relax ACM, 2010, pp. 513--522.

\bibitem{tong2011diversified}
H.~Tong, J.~He, Z.~Wen, R.~Konuru, and C.-Y. Lin, ``Diversified ranking on
  large graphs: an optimization viewpoint,'' in \emph{Proceedings of the 17th
  ACM SIGKDD international conference on Knowledge discovery and data
  mining}.\hskip 1em plus 0.5em minus 0.4em\relax ACM, 2011, pp. 1028--1036.

\bibitem{vondrak2013symmetry}
J.~Vondr{\'a}k, ``Symmetry and approximability of submodular maximization
  problems,'' \emph{SIAM Journal on Computing}, vol.~42, no.~1, pp. 265--304,
  2013.

\bibitem{wang2015friendbook}
Z.~Wang, J.~Liao, Q.~Cao, H.~Qi, and Z.~Wang, ``Friendbook: a semantic-based
  friend recommendation system for social networks,'' \emph{Mobile Computing,
  IEEE Transactions on}, vol.~14, no.~3, pp. 538--551, 2015.

\bibitem{wu2014fast}
Y.~Wu, R.~Jin, and X.~Zhang, ``Fast and unified local search for random walk
  based k-nearest-neighbor query in large graphs,'' in \emph{Proceedings of the
  2014 ACM SIGMOD international conference on Management of data}.\hskip 1em
  plus 0.5em minus 0.4em\relax ACM, 2014, pp. 1139--1150.

\bibitem{zhang2012evaluating}
C.~Zhang, L.~Shou, K.~Chen, G.~Chen, and Y.~Bei, ``Evaluating geo-social
  influence in location-based social networks,'' in \emph{Proceedings of the
  21st ACM international conference on Information and knowledge
  management}.\hskip 1em plus 0.5em minus 0.4em\relax ACM, 2012, pp.
  1442--1451.

\bibitem{zhang2015panther}
J.~Zhang, J.~Tang, C.~Ma, H.~Tong, Y.~Jing, and J.~Li, ``Panther: Fast top-k
  similarity search on large networks,'' in \emph{Proceedings of the 21th ACM
  SIGKDD International Conference on Knowledge Discovery and Data
  Mining}.\hskip 1em plus 0.5em minus 0.4em\relax ACM, 2015, pp. 1445--1454.

\bibitem{zhao2009p}
P.~Zhao, J.~Han, and Y.~Sun, ``P-rank: a comprehensive structural similarity
  measure over information networks,'' in \emph{Proceedings of the 18th ACM
  conference on Information and knowledge management}.\hskip 1em plus 0.5em
  minus 0.4em\relax ACM, 2009, pp. 553--562.

\end{thebibliography}

\end{document}